\documentclass{ieeeaccess}
\usepackage{cite}
\usepackage{amsmath,amssymb,amsfonts}
\usepackage{algorithmic}
\usepackage{graphicx}
\usepackage{textcomp}

\usepackage{url}
\usepackage{graphicx}
\usepackage{amsmath}
\usepackage{amssymb}
\usepackage{xcolor}
\usepackage{enumitem}
\usepackage{lipsum}
\usepackage{multirow}
\usepackage{array}
\usepackage{makecell}
\usepackage{diagbox}
\usepackage{caption}

\newcolumntype{L}[1]{>{\raggedright\let\newline\\\arraybackslash\hspace{0pt}}m{#1}}
\newcolumntype{C}[1]{>{\centering\let\newline\\\arraybackslash\hspace{0pt}}m{#1}}
\newcolumntype{R}[1]{>{\raggedleft\let\newline\\\arraybackslash\hspace{0pt}}m{#1}}
\newcolumntype{M}[1]{>{\centering\arraybackslash}m{#1}}
\newcolumntype{O}[1]{>{\raggedleft\arraybackslash}m{#1}}

\newlist{inlinelist}{enumerate*}{1}
\setlist[inlinelist]{label=(\arabic*)}
\newlist{inlinelistsub}{enumerate*}{1}
\setlist[inlinelistsub]{label=(\roman*)}

\usepackage[pagebackref=true,breaklinks=true,colorlinks,bookmarks=false]{hyperref}

\def\BibTeX{{\rm B\kern-.05em{\sc i\kern-.025em b}\kern-.08em
    T\kern-.1667em\lower.7ex\hbox{E}\kern-.125emX}}
\begin{document}
\history{Date of acceptance 2023-06-04.}
\doi{10.1109/ACCESS.2023.3284315}

\title{FPUS23: An Ultrasound Fetus Phantom Dataset with Deep Neural Network Evaluations for Fetus Orientations, Fetal Planes, and Anatomical Features}
\author{\uppercase{Bharath Srinivas Prabakaran\authorrefmark{1}, Paul Hamelmann\authorrefmark{2}, Erik Ostrowski}\authorrefmark{1}\IEEEmembership{Student Member, IEEE}, and \uppercase{Muhammad Shafique}\authorrefmark{3},
\IEEEmembership{Senior Member, IEEE}}
\address[1]{Institute of Computer Engineering, Technische Universit{\"a}t Wien (TU Wien), Austria (e-mail: \{bharath.prabakaran, erik.ostrowski\}@tuwien.ac.at)}
\address[2]{Philips Research, Eindhoven, The Netherlands (e-mail: paul.hamelmann@philips.com)}
\address[3]{Division of Engineering, New York University Abu Dhabi, United Arab Emirates (e-mail: muhammad.shafique@nyu.edu)}
\tfootnote{This work was partially supported by the Moore4Medical project funded by the ECSEL Joint Undertaking under grant number H2020-ECSEL-2019-IA-876190 and partially by the Doctoral College Resilient Embedded Systems, which is run jointly by TU Wien's Faculty of Informatics and UAS Technikum Wien. The authors also acknowledge TU Wien Bibliothek for financial support through its Open Access Funding Programme.}

\markboth
{Prabakaran \headeretal: FPUS23: An Ultrasound Fetus Phantom Dataset with Deep Neural Network Evaluations}
{Prabakaran \headeretal: FPUS23: An Ultrasound Fetus Phantom Dataset with Deep Neural Network Evaluations}

\corresp{Corresponding author: Bharath Srinivas Prabakaran (e-mail: bharath.prabakaran@tuwien.ac.at).}

\begin{abstract}
Ultrasound imaging is one of the most prominent technologies to evaluate the growth, progression, and overall health of a fetus during its gestation.
However, the interpretation of the data obtained from such studies is best left to expert physicians and technicians who are trained and well-versed in analyzing such images.
To improve the clinical workflow and potentially develop an at-home ultrasound-based fetal monitoring platform, we present a novel fetus phantom ultrasound dataset, FPUS23, which can be used to identify 
\begin{inlinelist}
    \item the correct diagnostic planes for estimating fetal biometric values, 
    \item fetus orientation, 
    \item their anatomical features, and 
    \item bounding boxes of the fetus phantom anatomies
\end{inlinelist}
at $23$ weeks gestation.
The entire dataset is composed of $15,728$ images, which are used to train four different Deep Neural Network models, built upon a ResNet34 backbone, for detecting aforementioned fetus features and use-cases.
We have also evaluated the models trained using our FPUS23 dataset, to show that the information learned by these models can be used to substantially increase the accuracy on real-world ultrasound fetus datasets. 
We make the FPUS23 dataset and the pre-trained models publicly accessible at \url{https://github.com/bharathprabakaran/FPUS23}, which will further facilitate future research on fetal ultrasound imaging.
\end{abstract}


\titlepgskip=-15pt

\maketitle


\section{Introduction}
\label{sec:Intro}

\begin{figure*}[t]
    \centering
    \includegraphics[width = \linewidth]{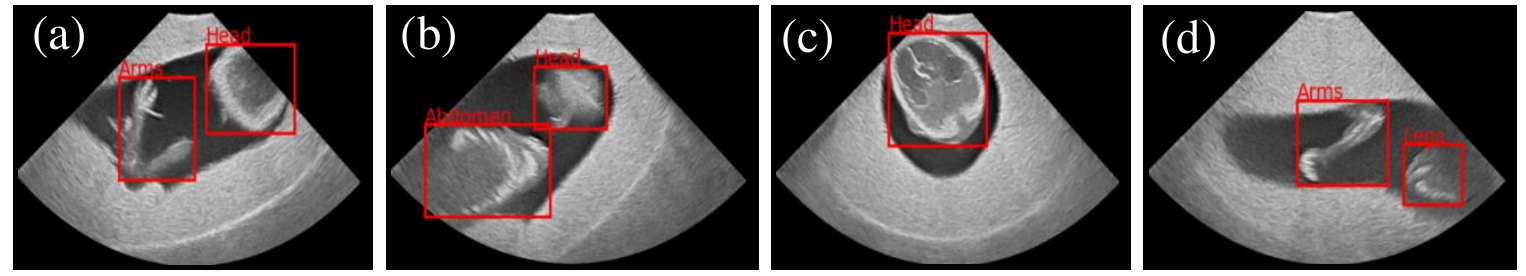}
    \caption{Identifying anatomical fetal features, such as limbs, which can enable the extraction of biometric parameters that determine the growth of the fetus. 
    Figures (a) - (d) illustrate the ability of an object detection model in detecting various fetus anatomies, across different views, with little fine-tuning on our FPUS23 dataset.}
    \label{fig:Mot}
\end{figure*}

\begin{figure*}[t]
    \centering
    \includegraphics[width = \linewidth]{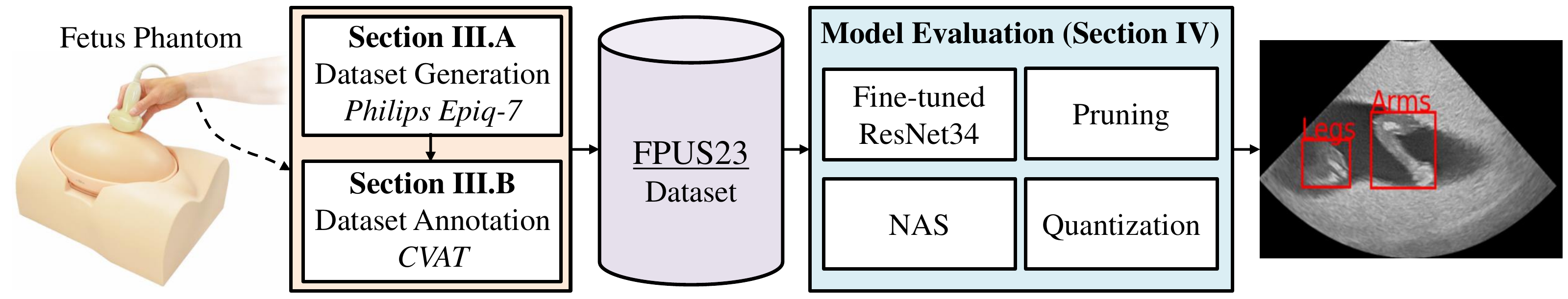}
    \caption{Overview of our methodology for generating and annotating the \textit{FPUS23} dataset.}
    \label{fig:Meth}
\end{figure*}

Ultrasound imaging techniques are used to create an image of organs and tissues inside the human body without the use of radiation, such as X-Rays, or expensive equipment, like Magnetic Resonance Imaging (MRI).
Ultrasound technologies are used in day-to-day healthcare clinics to efficiently diagnose diseases like COVID-19~\cite{born2020pocovid, ebadi2021covidx} and detect tumors~\cite{shi2016stacked}. 
Ultrasound is also widely used to monitor the development of an unborn human fetus to obtain information regarding its development and overall health.
An ultrasound examination is performed at various stages of the fetus' gestation to confirm the pregnancy and determine its location, condition, size, growth, orientation, gestational age, identify potential birth defects and complications, and many other factors relevant to the healthy development and delivery of the fetus.
However, the data obtained from such examinations are difficult to understand and require the expertise or training of sonographers or physicians to accurately interpret the data.
For instance, as depicted in an ultrasound image of a fetus at $23$ weeks gestation (fig.~\ref{fig:Mot}), identifying a fetus can be quite easy. 
However, identifying the orientation of the fetus and evaluating key biometric parameters, like the abdominal circumference or femur length, which are used to ascertain the gestational age of the fetus, requires the level of expertise that is currently offered only by trained sonographers and physicians.
This ``\textit{experience}'' can be learned and embedded within the deep learning models, which can be deployed in clinical use-cases as assistants to aid healthcare professionals in data interpretation.
These models can also be used to build an \textit{at-home portable ultrasound-based fetal monitoring platform} that can enable the user to understand the data by collating and interpreting the vital information. 

However, the development of such a deep learning model for analyzing fetal ultrasound data requires investigation of the following key research challenges:
\begin{inlinelist}
    \item ultrasound fetus data fall under the category of healthcare information that is heavily protected by regulatory requirements regarding their generation, storage, usage, etc. to ensure patient privacy;
    \item due to these regulations, there are very few openly accessible fetal ultrasound datasets, which can be used to develop such clinical assistants and at-home monitoring platforms;
    \item even the state-of-the-art datasets that are accessible are not properly annotated with the relevant information in order to be able to train DNN models, which can be used to infer relevant fetus anatomy information; and 
    \item the existing datasets are not large enough to enable the DNN model to learn the required features efficiently, despite the use of existing transfer learning approaches.
\end{inlinelist}

\begin{figure*}[t]
    \centering
    \includegraphics[width = \linewidth]{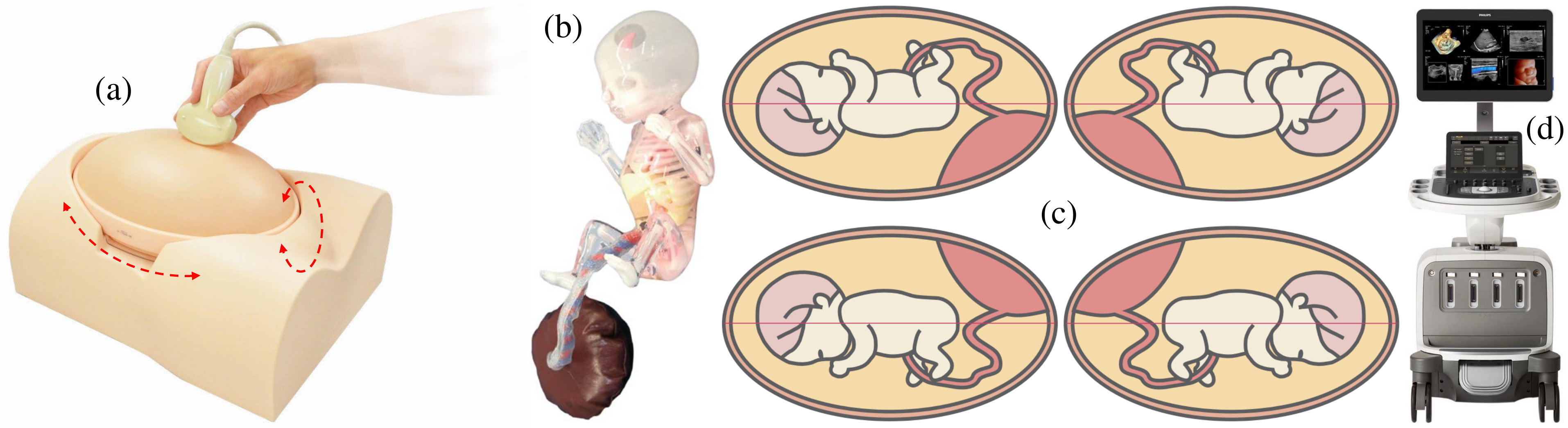}
    \caption{Overview of (a) the phantom abdomen in the mother body torso and its possible rotations; (b) the fetus phantom placed in the abdomen; (c) the four possible fetus orientations; (d) the Philips Epiq-$7$ ultrasound system (adapted from~\cite{EPIQ}~and~\cite{FetusPhantom}).}
    \label{fig:FetPhan}
\end{figure*}

To address these research challenges, we build the \textit{FPUS23} dataset (see fig.~\ref{fig:Meth}) by:
\begin{inlinelist}
    \item using a fetus phantom at $23$ weeks gestation, instead of actual human fetuses, thereby circumventing the regulations associated with healthcare data;
    \item not generating or using any healthcare data in our dataset -- FPUS23 will be openly accessible to further facilitate future research and advancements in this domain;
    \item properly labeling and annotating the dataset with the help of scientists with experience in fetal ultrasound imaging to generate a dataset that can be used to identify
    \begin{inlinelistsub}
        \item diagnostic planes for extracting fetal biometric parameters,
        \item fetus orientation,
        \item fetus anatomies, and
        \item bounding boxes of the fetus anatomies;
    \end{inlinelistsub}
    \item building a dataset with 15,728 ultrasound image samples that can be used to learn the required information;
    \item extensive evaluation of our datasets with appropriate transfer learning approaches, including model compression techniques, as illustrated in Section~\ref{sec:ER}. %
\end{inlinelist}


\section{Related Work}
\label{sec:RW}

There is an abundance of ultrasound datasets for various use-cases, which can be used to generate DNN-based models for classification and segmentation. 
For instance, the breast ultrasound image dataset presented by Al-Dhabyani~\textit{et~al.}~\cite{al2020dataset}, which is composed of normal, benign, and malignant images that can be used to train to a model to act as a classifier.
Similarly, the POCUS dataset, presented by Born~\textit{et~al.}~\cite{born2020pocovid}, and the COVIDX-US dataset, by Ebadi~\textit{et~al.}~\cite{ebadi2021covidx}, are openly accessible for building DNN-based clinical assistants that can aid in the analytics and diagnosis of COVID-19.
Leclerc~\textit{et~al.}~\cite{leclerc2019deep} presented a cardiac ultrasound electrocardiography dataset containing image sequences with two and four-chamber views of the heart of $500$ patients.
Likewise, there are a wide number of ultrasound datasets for diagnosing and analyzing several internal body organs. 

\textbf{Fetal Ultrasound:}
Deep learning has also been explored for fetal ultrasound imaging, albeit not as widely or comprehensively.
\cite{valanarasu2020learning} proposed a multi-scale self-attention generator that can be used to automatically generate ultrasound images from various segmentation masks, which can then be used for fetal brain segmentation and analysis.
\cite{patra2017learning} proposed the use of deep learning to automatically analyze the fetal heart by encoding and translating Spatio-temporal information in order to classify amongst three different fetal heart planes.
\cite{komatsu2021detection} proposed a CNN-based classifier that can be used to detect cardiac abnormalities in fetal ultrasound images.
\cite{qu2019deep} have proposed the use of CNNs to detect the six standard fetal brain planes on a proprietary dataset containing $30,000$ $2$D fetal ultrasound images gathered between $16$ and $34$ weeks gestation, to achieve $91\%$ accuracy.
\cite{van2019automated} presented two DNN models that are used to detect the ideal frame for the fetal head, followed by its segmentation, which is used to measure the head circumference, a key biometric parameter.
\cite{sobhaninia2019fetal} proposed an improved multi-task learning network that improves the segmentation capabilities of the model when compared to~\cite{van2019automated}.
Although ultrasound examinations of a fetus are very common, there are very few openly accessible datasets for researchers to build DNN-based models that can aid in the analytics of fetal ultrasound images.
\cite{van2018automated} presented a dataset of fetal ultrasound images with annotations regarding the head circumference, as shown in~\cite{van2019automated,sobhaninia2019fetal}.
\cite{burgos2020evaluation} presented a fetal ultrasound dataset with over $12,400$ images from $1,792$ patients, which were categorized into six  classes containing the anatomical planes.
Most of the other works in this category primarily work on proprietary datasets, which are not accessible for analysis and evaluation.


\section{FPUS23: The Fetal Ultrasound Dataset}
\label{sec:USFP}

\subsection{Data Collection}

The required ultrasound fetal data was generated and collected using the ``US-7 SPACE FAN-ST'' fetus phantom~\cite{FetusPhantom}, which has been typically used to train sonographers to assess the development and condition of the fetus.
Figs.~\ref{fig:FetPhan}(a)~and~\ref{fig:FetPhan}(b) illustrate an overview of the oval-shaped phantom abdomen, which mimics the uterus containing a fetus at $23$ weeks gestation, and the life-size fetus demonstration model that is placed inside the phantom abdomen, respectively.
We propose to use a $23$-week old phantom as a mid-pregnancy scan is typically performed around this time to check for fetus anomalies.
The phantom abdomen can be rotated into four different positions to change the orientation and presentation (cephalic or breech) of the fetus phantom (see fig.~\ref{fig:FetPhan}(c)).
The fetus model includes full skeletal structure and key organic features that can be observed and used to train the sonographer to assess the fetus' anatomy (like head, arms legs, abdomen) and internal body organs (like brain, skull, spine, cardiac chambers, stomach, kidney, blood vasculature, etc.).
The biometric parameters of the fetus can also be measured/learned using an ultrasound of the fetus phantom at the appropriate positions or the correct \textit{diagnostic planes}.
Besides the aforementioned features, the quantity of amniotic fluid, any potential abnormalities, location of the placenta, fetal posture, etc. can also be learned with the help of this model.

\begin{figure*}[t]
    \centering
    \includegraphics[width = 0.95\linewidth]{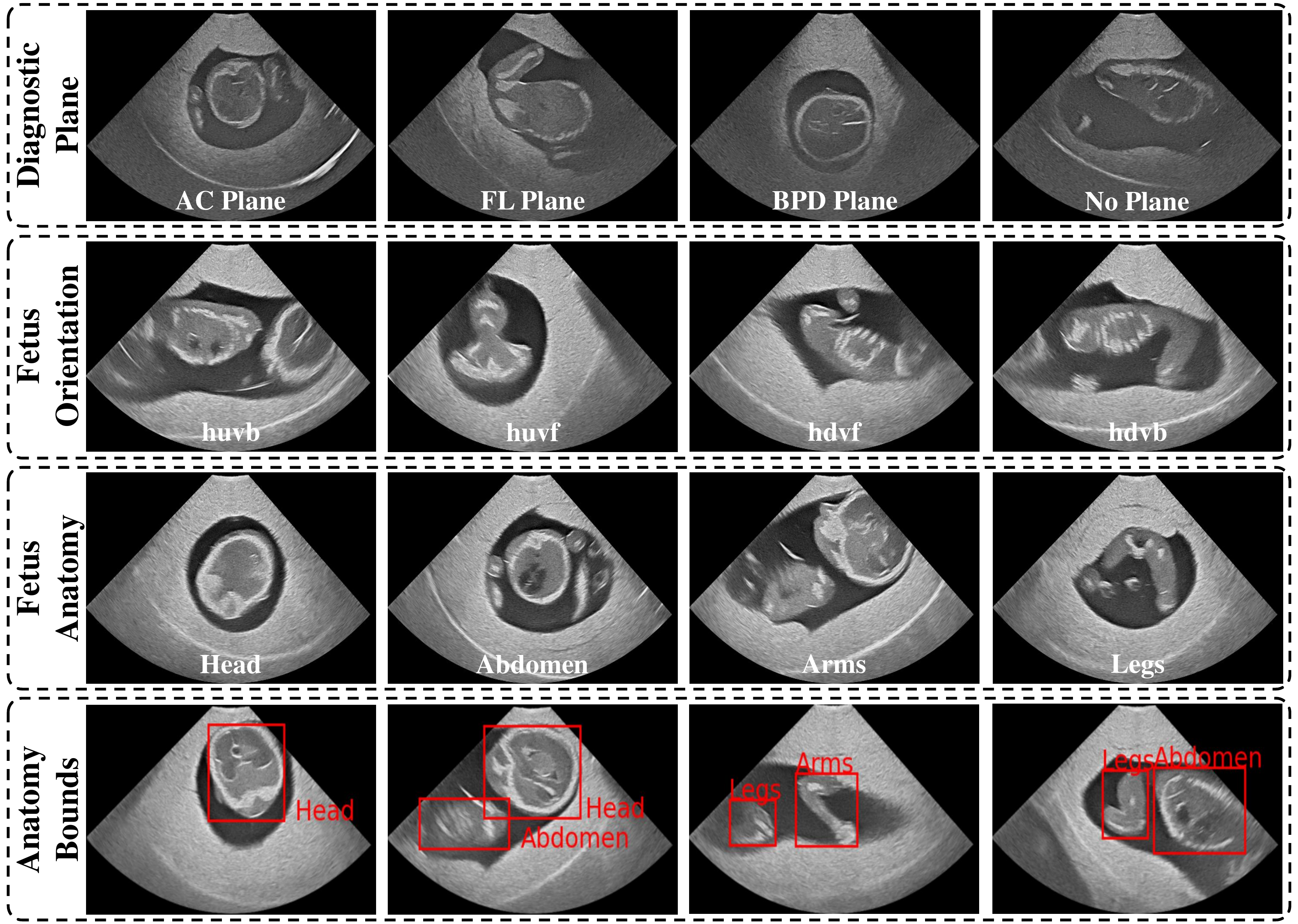}
    \caption{Super-labels of the \textit{FPUS23} dataset and corresponding samples in each class.}
    \label{fig:FPUS23Ann}
\end{figure*}

We use the X$6$-$1$ xMATRIX array transducer~\cite{MatTrans}, which is interfaced with the Philips Epiq-$7$ system~\cite{EPIQ} to collect and process the data to generate the final ultrasound image (see fig.~\ref{fig:FetPhan}(d)).
The Anatomically Intelligent Ultrasound (AIUS) imaging technology deploys advanced organ modeling and imaging techniques to generate a two-dimensional image of the fetus phantom using the default settings for the ``OB Fetal Echo'' imaging option.
The imaging depth was set to $12$cm and captured at a $23$Hz frame rate.
Sufficient ultrasound gel is applied on the phantom abdomen to ensure acoustic coupling with the probe, thereby reducing acoustic impedance, and enabling clear imaging.
We executed two protocols to collect the images used in the \textit{FPUS23} dataset:

\begin{enumerate}[label=(\arabic*),leftmargin=*]
    \item \noindent\textbf{Protocol-I}: The probe is placed on the phantom abdomen surface and navigated to the \textit{correct diagnostic planes} that can be used for the measurement of the three primary biometric parameters of the fetus, namely the transvetricular plane, which is used to obtain the brain's Biparietal Diameter~(BPD), abdominal standard plane, which is used to estimate the Abdominal Circumference~(AC), and the femur standard plane, which is used to estimate the fetus' Femur Length~(FL).
    The correct diagnostic planes were identified using the clinical protocols discussed by Salomon~\textit{et~al.}~\cite{salomon2011practice} and Bethune~\textit{et~al.}~\cite{bethune2013pictorial}.
    To further enrich the dataset, after the acquisition of several frames at the correct diagnostic plane, we tilt, rotate, or traverse the ultrasound probe in random directions to collect more information.
    \item \textbf{Protocol-II}: The focus of this protocol is to obtain images capturing the anatomies of the fetus phantom in the generated images.
    We do this by navigating the probe to obtain the head, abdomen, arms, and legs, individually or combined, in the picture and move the probe in different directions to obtain a heterogeneous set of images capturing the fetal anatomies.
    Furthermore, the phantom abdomen was also rotated and placed in the four possible orientations [head up (hu) or down (hd), view front (vf) or back (vb)], when collecting the ultrasound data, to potentially mimic the real-life behavior of fetus orientation and presentation (see fig.~\ref{fig:FetPhan}(c)).
    Additionally, the probe orientation was also changed between horizontal and vertical, with respect to the abdomen, when the data was collected to enhance the dataset with more information.
\end{enumerate}



\subsection{Annotation}

The data streams obtained by the Philips Epiq-$7$ ultrasound system are converted to PNG image sequences, of dimension $664$x$388$, using custom in-house software for easier labeling, annotating, and processing.
The stored PNG files are annotated using a customized version of the Computer Vision Annotation Tool -- CVAT~\cite{boris_sekachev_2020_4009388}, which was primarily opted for its ease-of-use and wide-range features.
Each acquired ultrasound frame was subsequently annotated by scientists with experience in fetal ultrasound imaging. 

The sequences obtained using \textit{Protocol-I} are labeled as a correct diagnostic plane for one of the three biometric parameters (BPD, AC, FL) or as a non-diagnostic plane.
Since the number of data samples obtained for each of the diagnostic planes is quite smaller than the non-diagnostic plane output class, the samples were augmented in each of the other three output classes to ensure equal representation of data across all classes.
The data obtained using \textit{Protocol-II} is, first, labeled with fetus orientations, namely huvf, huvb, hdvf, or hdvb, as discussed earlier, based on the position of the phantom abdomen when the scans are made.
Next, the images are tagged with the anatomies present in the image, such as heads, arms, legs, and abdomen.
The images are subsequently exhaustively annotated with boxes representing the respective anatomies to determine their bounds and potentially estimate biometric parameters later, such as femur length, at the correct diagnostic plane.
Images that do not contain any vital and/or relevant information regarding the fetus are not labeled or annotated.
These finalized labels and annotations, for each valid image in the dataset, are extracted as an XML file, which can be used to train various deep learning models as required.

Table~\ref{tab:FPUS23BD} depicts an overview of our \textit{FPUS23} dataset and the number of input samples present in each class for each of the four different super-labels:
\begin{inlinelist}
    \item Diagnostic Plane,
    \item Fetus Orientation,
    \item Fetus Anatomy, and
    \item Anatomy Bounds, using box annotation.
\end{inlinelist}
The dataset is split in the ratio of $8:1:1$, with respect to training, validation, and testing, respectively, for the first three cases.
We split the anatomy bounds data and use $80\%$ of it for training and $20\%$ for validating the model. 
Fig.~\ref{fig:FPUS23Ann} depicts a few sample images and their corresponding annotations for the four different super-labels of our \textit{FPUS23} dataset.

\begin{table}[t]
\centering
\caption{Breakdown of the number of labeled data samples in each class of our \textit{FPUS23} dataset; Of the $15,728$ images, not all contain relevant information for labeling -- images obtained during probe movements may not accurately depict anatomies.}
\begin{tabular}{cccccc}
\multicolumn{1}{c|}{\multirow{4}{*}{\rotatebox[origin=c]{90}{\shortstack{Diagnostic\\ Plane}}}}  & \multicolumn{4}{c}{Total}                  \\ \cline{2-5} 
\multicolumn{1}{c|}{}                                  & \multicolumn{4}{c}{$5,265$}                \\
\multicolumn{1}{c|}{}                                  & AC Plane & BPD Plane & FL Plane & No Plane \\ \cline{2-5} 
\multicolumn{1}{c|}{}                                  & $1,386$  & $1,280$   & $1,281$  & $1,318$  \\ 
                                  & & & & \\ 
\multicolumn{1}{c|}{\multirow{4}{*}{\rotatebox[origin=c]{90}{\shortstack{Fetus\\ Orientation}}}} & \multicolumn{4}{c}{Total}                  \\ \cline{2-5} 
\multicolumn{1}{c|}{}                                  & \multicolumn{4}{c}{$15,728$}               \\
\multicolumn{1}{c|}{}                                  & hdvb     & hdvf      & huvb     & huvf     \\ \cline{2-5} 
\multicolumn{1}{c|}{}                                  & $3,757$  & $3,235$   & $3,980$  & $4,756$  \\
                                  & & & & \\ 
\multicolumn{1}{c|}{\multirow{4}{*}{\rotatebox[origin=c]{90}{\shortstack{Fetus\\ Anatomy}}}}     & \multicolumn{4}{c}{Total}                  \\ \cline{2-5} 
\multicolumn{1}{c|}{}                                  & \multicolumn{4}{c}{$9,317$}                \\
\multicolumn{1}{c|}{}                                  & Head     & Arms      & Legs     & Abdomen  \\ \cline{2-5} 
\multicolumn{1}{c|}{}                                  & $3,003$  & $1,629$   & $2,159$  & $2,526$  \\
                                  & & & & \\ 
\multicolumn{1}{c|}{\multirow{4}{*}{\rotatebox[origin=c]{90}{\shortstack{Anatomy\\ Bounds}}}}        & \multicolumn{4}{c}{Total}                  \\ \cline{2-5} 
\multicolumn{1}{c|}{}                                  & \multicolumn{4}{c}{$9,455$}                \\
\multicolumn{1}{c|}{}                                  & Head     & Arms      & Legs     & Abdomen  \\ \cline{2-5} 
\multicolumn{1}{c|}{}                                  & $4,370$    & $4,853$   & $4,572$  & $6,435$
\end{tabular}
\label{tab:FPUS23BD}
\end{table}


\section{Experiments and Results}
\label{sec:ER}

\subsection{Experimental Setup}

The experimental evaluations illustrated in this section, which depict the efficacy of the DNN models trained using our FPUS23 dataset, are primarily completed on a CentOS $7.9$ Operating System running on an Intel Core i$7$-$8700$ CPU with $16$GB RAM and $2$ Nvidia GeForce GTX $1080$ Ti GPUs.
Our scripts were executed with the following software versions: CUDA $11.5$, Pytorch $3.7.4.3$, torchvision $0.11.1$, and Pytorch-lightning $1.5.1$.
We use a ResNet34~\cite{he2016deep} DNN model, pre-trained using the ImageNet~\cite{deng2009imagenet}, and retrain it with our dataset for $15$ epochs using the cross-entropy loss function.
The initial and final layers of the ResNet34 architecture were adapted to accommodate the custom input data dimensions and output classes, respectively.
We use a modified Faster-RCNN~\cite{ren2015faster} with our ResNet34 backbone to build the model used for determining the anatomy bounds of the fetus in our dataset.
The learning rate for all models was set to $0.001$ using the Adam Optimizer with a step size of $20$ and $\gamma=0.1$.

We use the traditional metrics of accuracy, precision, and recall to determine the efficacy of the deep learning models illustrated in this section.
Accuracy is the ratio of the total number of inputs accurately predicted with respect to the total number of predictions, which is a primary evaluation metric for classification systems.
For the anatomy bounds, we use the mean Average Precision (\textit{mAP}; IoU=[$.50$:$.05$:$.95$]), which is the ability of the model to not label a negative sample as positive, and mean Average Recall (\textit{mAR}; IoU=[$.50$:$.05$:$.95$]), which denotes the ability of the model to identify all positive instances of each class:
\begin{equation}
mAP = \frac{1}{N} \frac{\sum_{i=1}^N p_{i,i}}{\sum_{i=0}^N \sum_{j=1}^N p_{i,i} + p_{i,j}}
\end{equation}
\begin{equation}
mAR = \frac{1}{N} \frac{\sum_{i=1}^N p_{i,i}}{\sum_{i=0}^N \sum_{j=1}^N p_{i,i} + p_{j,i}}, i,j = 1, ..., N
\end{equation}
where $N$ denotes the total number of output classes, $p_{i,i}$ the number of pixels classified as class $i$ and labeled as class $i$, and $p_{i,j}$, $p_{j,i}$ are the number of pixels classified as class $i$ and labeled as class $j$ and vice-versa.
We also evaluate the model's F1-score, which is the harmonic mean of the model's precision and recall.

\begin{table}[t]
\centering
\caption{Preliminary quality evaluations (accuracy and F1-score) and hardware requirements (Flops and memory) of the modified ResNet34 model trained using the FPUS23 dataset.}
\begin{tabular}{c|cccc}
\multicolumn{1}{c|}{} & \multirow{2}{*}{Accuracy}     & \multirow{2}{*}{\shortstack{F1-\\Score}}       & \multirow{2}{*}{\shortstack{No. of\\Flops}}                  & \multirow{2}{*}{\shortstack{Memory\\(MB)}}\\
& & & & \\\cline{1-5}
\multirow{2}{*}{\shortstack{Diagnostic\\ Plane}}     & \multirow{2}{*}{$98.29\%$}          & \multirow{2}{*}{$.9833$}         & \multirow{6}{*}{$3.67$G} & \multirow{6}{*}{$157.57$} \\
& & & & \\
\multirow{2}{*}{\shortstack{Fetus\\ Orientation}}    & \multirow{2}{*}{$99.80\%$}          & \multirow{2}{*}{$.9936$}         &                        &                         \\
& & & & \\
\multirow{2}{*}{\shortstack{Fetus\\ Anatomy}}        & \multirow{2}{*}{$99.46\%$}          & \multirow{2}{*}{$.9989$}         &                        &                         \\
& & & & \\
\multirow{2}{*}{\shortstack{Anatomy\\ Bounds}} & mAP & mAR & \multirow{2}{*}{$64.4$G} & \multirow{2}{*}{$357.20$}                       \\\cline{2-3}
 & $98.60\%$ & $84.40\%$
\end{tabular}
\label{tab:ResNet34}
\end{table}

\begin{table*}[t]
\centering
\caption{Exhaustive evaluations of the compressed baseline ResNet34 model.}
{\fontsize{10pt}{10pt}\selectfont
\begin{tabular}{rcccccccc}
                     & \multicolumn{8}{c}{Pruning}                                                                                                                                                           \\
                     & \multicolumn{2}{c}{$0\%$}                   & \multicolumn{2}{c}{$30\%$}                  & \multicolumn{2}{c}{$50\%$}                  & \multicolumn{2}{c}{$70\%$}                  \\ \cline{2-9}
                     & \\
                     & \multicolumn{8}{c}{Quantization}                                                                                                                                                      \\
                     & FP$32$               & INT$8$               & FP$32$               & INT$8$               & FP$32$               & INT$8$               & FP$32$               & INT$8$               \\ \cline{2-9} 
                     & \\
No. of Flops         & $3.67$G              & -                    & $2.58$G              & -                    & $1.85$G              & -                    & $1.12$G              & -                    \\
Memory (MB)            & $157.57$             & $39.39$              & $125.82$             & $31.45$              & $104.65$             & $26.16$              & $83.48$              & $20.87$              \\
& \\
                     & \multicolumn{8}{c}{Diagnostic Planes}                                                                                                                                                 \\ \cline{2-9}
Accuracy             & $98.29\%$            & $97.34\%$            & $98.10\%$            & $98.29\%$            & $98.10\%$            & $97.91\%$            & $97.72\%$            & $\mathbf{99.25\%}$   \\
F1-Score             & $.9833$              & $.9759$              & $.9818$              & $.9832$              & $.9813$              & $.9799$              & $.9799$              & $\mathbf{.9834}$     \\
& \\
                     & \multicolumn{8}{c}{Fetus Orientation}                                                                                                                                                 \\ \cline{2-9} 
Accuracy             & $99.80\%$            & $99.73\%$            & $99.49\%$            & $\mathbf{99.93\%}$   & $99.55\%$            & $99.87\%$            & $99.68\%$            & $\mathbf{99.93\%}$   \\
F1-Score             & $.9936$              & $.9992$              & $.9949$              & $\mathbf{.9994}$     & $.9956$              & $.9988$              & $.9968$              & $\mathbf{.9994}$     \\
& \\
                     & \multicolumn{8}{c}{Fetus Anatomy}                                                                                                                                                     \\ \cline{2-9} 
Accuracy             & $99.46\%$            & $99.46\%$            & $99.57\%$            & $\mathbf{99.89\%}$   & $99.78\%$            & $\mathbf{99.89\%}$   & $99.67\%$            & $99.78\%$            \\
F1-Score             & $\mathbf{.9989}$     & $.9749$              & $.9952$              & $.9987$              & $.9978$              & $.9987$              & $.9968$              & $.9978$              \\
& \\
                     & \multicolumn{8}{c}{Anatomy Bounds - Faster-RCNN with ResNet34 backbone}                                                                                                               \\ \cline{2-9} 
mAP                  & $98.60\%$            & $98.60\%$            & $98.00\%$            & $98.00\%$            & $\mathbf{98.80\%}$   & $97.80\%$            & $98.40\%$            & $98.00\%$            \\
mAR                  & $84.40\%$            & $81.20\%$            & $\mathbf{87.90\%}$   & $78.50\%$            & $85.90\%$            & $78.60\%$            & $86.00\%$            & $79.80\%$            \\
No. of Flops         & $64.40$G             & -                    & $45.09$G             & -                    & $32.21$G             & -                    & $19.33$G             & -                    \\
Memory (MB)            & $357.20$             & $239.02$             & $250.04$             & $155.67$             & $178.60$             & $100.11$             & $107.16$             & $44.55$              \\
\end{tabular}}
\label{tab:NASEval1}
\end{table*}

\begin{table*}[t]
\centering
\caption{Evaluations of the DNN model obtained through exhaustive neural architecture search; As expected, a smaller model achieves the same or similar output quality as the baseline while reducing the hardware requirements by up to $17\times$.}
{\fontsize{10pt}{10pt}\selectfont
\begin{tabular}{rcccccccc}
                     & \multicolumn{8}{c}{Pruning}                                                                                                                                                           \\
                     & \multicolumn{2}{c}{$0\%$}                   & \multicolumn{2}{c}{$30\%$}                  & \multicolumn{2}{c}{$50\%$}                  & \multicolumn{2}{c}{$70\%$}                  \\ \cline{2-9} 
                     & \\
                     & \multicolumn{8}{c}{Quantization}                                                                                                                                                      \\
                     & FP$32$               & INT$8$               & FP$32$               & INT$8$               & FP$32$               & INT$8$               & FP$32$               & INT$8$               \\ \cline{2-9} 
                     & \\
No. of Flops         & $1.42$G              & -                    & $1.00$G              & -                    & $0.73$G              & -                    & $0.45$G              & -                    \\
Memory (MB)            & $41.26$              & $10.31$              & $37.03$              & $9.25$               & $34.21$              & $8.55$               & $31.38$              & $7.84$               \\
\multicolumn{1}{l}{} & \multicolumn{1}{l}{} & \multicolumn{1}{l}{} & \multicolumn{1}{l}{} & \multicolumn{1}{l}{} & \multicolumn{1}{l}{} & \multicolumn{1}{l}{} & \multicolumn{1}{l}{} & \multicolumn{1}{l}{} \\
                     & \multicolumn{8}{c}{Diagnostic Planes}                                                                                                                                                 \\ \cline{2-9}
Accuracy             & $99.24\%$            & $98.29\%$            & $\mathbf{99.62\%}$   & $98.67\%$            & $99.24\%$            & $99.43\%$            & $99.24\%$            & $97.72\%$            \\
F1-Score             & $.9930$              & $.9821$              & $\mathbf{.9965}$     & $.9877$              & $.9930$              & $.9947$              & $.9930$              & $.9789$              \\
\multicolumn{1}{l}{} & \multicolumn{1}{l}{} & \multicolumn{1}{l}{} & \multicolumn{1}{l}{} & \multicolumn{1}{l}{} & \multicolumn{1}{l}{} & \multicolumn{1}{l}{} & \multicolumn{1}{l}{} & \multicolumn{1}{l}{} \\
                     & \multicolumn{8}{c}{Fetus Orientation}                                                                                                                                                 \\ \cline{2-9} 
Accuracy             & $99.57\%$            & $99.61\%$            & $\mathbf{99.83\%}$   & $98.60\%$            & $99.53\%$            & $99.80\%$            & $99.70\%$            & $99.80\%$            \\
F1-Score             & $.9980$              & $.9962$              & $.9976$              & $.9861$              & $.9935$              & $\mathbf{.9982}$     & $.9957$              & $\mathbf{.9982}$     \\
\multicolumn{1}{l}{} & \multicolumn{1}{l}{} & \multicolumn{1}{l}{} & \multicolumn{1}{l}{} & \multicolumn{1}{l}{} & \multicolumn{1}{l}{} & \multicolumn{1}{l}{} & \multicolumn{1}{l}{} & \multicolumn{1}{l}{} \\
                     & \multicolumn{8}{c}{Fetus Anatomy}                                                                                                                                                     \\ \cline{2-9} 
Accuracy             & $\mathbf{99.89\%}$   & $99.78\%$            & $\mathbf{99.89\%}$   & $\mathbf{99.89\%}$   & $\mathbf{99.89\%}$   & $99.67\%$            & $99.78\%$            & $99.46\%$            \\
F1-Score             & $\mathbf{.9987}$     & $.9975$              & $\mathbf{.9987}$     & $\mathbf{.9987}$     & $\mathbf{.9987}$     & $.9969$              & $.9975$              & $.9934$              \\
\multicolumn{1}{l}{} & \multicolumn{1}{l}{} & \multicolumn{1}{l}{} & \multicolumn{1}{l}{} & \multicolumn{1}{l}{} & \multicolumn{1}{l}{} & \multicolumn{1}{l}{} & \multicolumn{1}{l}{} & \multicolumn{1}{l}{} \\
                     & \multicolumn{8}{c}{Anatomy Bounds - Faster-RCNN with ResNet10 backbone}                                                                                                               \\ \cline{2-9} 
mAP                  & $\mathbf{97.90\%}$   & $\mathbf{97.90\%}$   & $97.30\%$            & $97.50\%$            & $97.20\%$            & $97.60\%$            & $97.20\%$            & $97.30\%$            \\
mAR                  & $\mathbf{81.30\%}$   & $78.90\%$            & $75.80\%$            & $75.30\%$            & $76.20\%$            & $75.70\%$            & $76.90\%$            & $75.00\%$            \\
No. of Flops         & $28.18$G             & -                    & $19.74$G             & -                    & $14.11$G             & -                    & $8.46$G              & -                    \\
Memory (MB)            & $290.84$             & $194.61$             & $203.59$             & $126.74$             & $145.42$             & $81.51$              & $87.25$              & $36.27$              \\
\multicolumn{1}{l}{} & \multicolumn{1}{l}{} & \multicolumn{1}{l}{} & \multicolumn{1}{l}{} & \multicolumn{1}{l}{} & \multicolumn{1}{l}{} & \multicolumn{1}{l}{} & \multicolumn{1}{l}{} & \multicolumn{1}{l}{} 
\end{tabular}}
\label{tab:NASEval2}
\end{table*}

\subsection{Baseline Model}

To illustrate the effectiveness of the \textit{FPUS23} dataset, we retrain a modified version of the ResNet34 architecture, which we consider as the baseline, to illustrate the capability of the network to learn relevant information regarding the classification of labels and detecting fetal anatomies.
Accuracy and F1-score are the two metrics used to determine the quality of the model, whereas the number of floating-point operations (Flops) and memory (MB) are relevant to estimate the hardware and resource requirements of the baseline model and determine its deployability in edge devices.
The results of these experiments are illustrated in Table~\ref{tab:ResNet34}. 
Achieving $\sim99\%$ accuracy in the classification of diagnostic planes, fetus orientation, and fetus anatomy gives us the understanding that the models are able to learn the features quite well.
Similarly, the modified Faster-RCNN model, embedded with our ResNet34 backbone, is able to detect fetal anatomies at significantly high precision.
Note, the significantly high quality of the models can also be attributed to their over-parameterization, which implies that smaller networks, achieving similar output accuracy, can be obtained with Neural Architecture Search (NAS)~\cite{elsken2019neural} and model compression techniques \cite{cheng2017survey} (see Section~\ref{subsec:MC}).

Moreover, the use of an inverted probe during an ultrasound exam leads to the generation of an inverted image.
To design a robust network model that can extract features and information from the inverted image to ensure correct classification, relevant image samples need to be collected, with the probe inverted, and included in the dataset during the training stage. 
However, this information can also be added to the model by flipping the collected images along the y-axis, which mimics probe-inverted images, and adding them to the original dataset before training.

\subsection{Model Compression}
\label{subsec:MC}
To further reduce the model's hardware requirements, we use compression techniques like pruning and quantization.
We have implemented the technique proposed by Han~\textit{et~al.}~\cite{han2015deep}, which proposes to eliminate the smallest $x\%$ of total weights and associated connections from the network, followed by a network retraining stage, wherein the model relearns the information on the reduced set of available parameters, to potentially achieve similar output quality as the original model.
We analyze the quality and hardware require-ments of the models that are $30\%$, $50\%$, and $70\%$ pruned.
The pruned networks are subsequently quantized to $8$-bit integer (INT$8$) precision, using the quantization-aware training strategy presented by Khudia~\textit{et~al.} \cite{khudia2021fbgemm}, to further reduce the model's size and improve its computational performance; INT$8$ computations are several orders of magnitude faster than $32$-bit floating-point (FP$32$) operations \cite{sze2017efficient}.
Similar to the regularizing effect illustrated in~\cite{marchisio2018prunet}, compression of the baseline models led to potential scenarios where the compressed models outperform the original.
This regularizing effect is especially prominent when the ResNet34 classifiers are compressed, due to their heavy over-parameterization.
Table~\ref{tab:NASEval1} illustrates the quality evaluations and the hardware requirements of these models when trained on \textit{FPUS23}.

Plenty of research works on automated NAS have demonstrated their efficacy in reducing the number of parameters and computations for achieving similar quality results compared to over-parameterized architectures that are typically designed by hand.
Towards this, we first investigate the effectiveness of two state-of-the-art automated NAS approaches presented by Fang~\textit{et~al.}~\cite{fang2020densely} and Wang~\textit{et~al.}~\cite{wang2020apq} in reducing the number of Flops and memory while retaining the output quality.
Both these approaches yield a residual DNN with $10$ intermediate feature extraction layers to achieve an output quality similar to that of the baseline.
However, while exhaustively generating and exploring smaller networks from scratch, we generated a residual DNN with $8$ layers, instead of the $10$ proposed by~\cite{fang2020densely,wang2020apq}, which offers similar output quality while requiring a fewer number of parameters.
We use this ResNet8 architecture to build models for classifying the diagnostic planes, fetus orientation, and fetus anatomy.
However, we have observed that the quality of the Faster-RCNN model, built using a ResNet8, is substantially lower as opposed to the model using a ResNet10 backbone.
Therefore, the Faster-RCNN model is built using a ResNet10 backbone instead.
From the results, it is quite evident that the ResNet8 model achieves a similar output quality as the original baseline models, even when both of them are compressed using pruning and/or quantization. 
For instance, a fetus anatomy classifier built using the ResNet8 architecture, which has been $30\%$ pruned and INT$8$ quantized, achieves an output quality greater than the baseline ResNet34 model, while requiring less than $10$MB in memory, making it ideal for deployment on resource-constrained processing platforms.
The Faster-RCNN built using the ResNet10 backbone also achieves similar quality to the baseline model while substantially reducing the hardware requirements.
Table~\ref{tab:NASEval2} illustrates the quality evaluations and the hardware requirements of the NAS models when trained on the \textit{FPUS23} dataset.

\begin{figure*}[t]
    \centering
    \includegraphics[width = \linewidth]{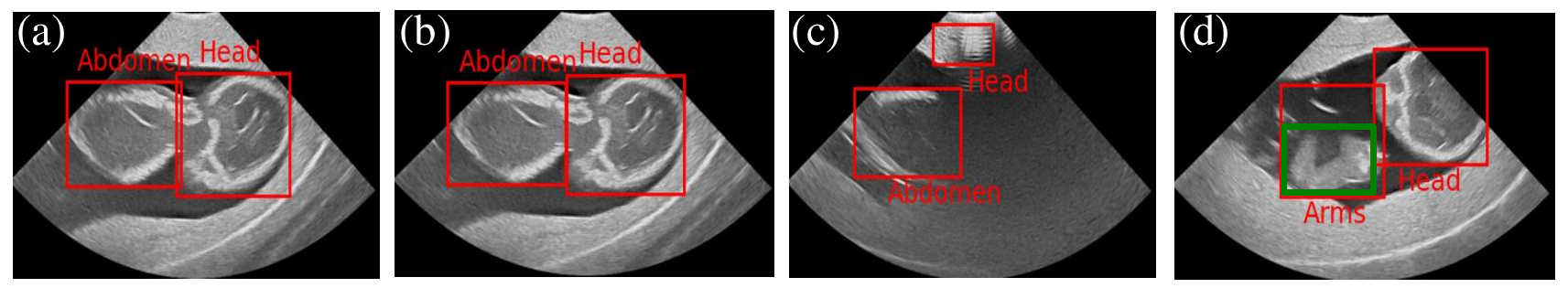}
    \caption{Evaluation of Faster-RCNN trained on \textit{FPUS23}; comparison between (a) ground truth and (b) prediction; (c) detecting anatomies incorrectly during probe navigation; (d) misestimating dimensions of fetus anatomy.}
    \label{fig:ODEval}
\end{figure*}

\subsection{Evaluation on State-of-the-Art Real-World Dataset}

To further demonstrate the applicability of our dataset in real-world fetal ultrasound use-cases, we fine-tune the models trained using \textit{FPUS23} on the training set and evaluate them on the test set presented by~\cite{burgos2020evaluation}. 
We train two models: first, we train the baseline model on the \textit{FPUS23} dataset, before retraining on the real-world dataset, and in the second, we train the model directly on the real-world dataset. 
Model-$1$ achieves $91.92\%$ accuracy in detecting the anatomical planes after training for just $1$ epoch, whereas Model-$2$ achieves the same accuracy only after training for more than $16$ epochs.
Therefore, the models can learn relevant features regarding fetal ultrasounds from the \textit{FPUS23} dataset before being deployed for other fetal ultrasound datasets with very little fine-tuning.
To re-emphasize this, we perform a small analysis on the networks under consideration. 
We consider the weights of the FPUS23-trained model before and after fine-tuning on the real-world dataset. 
Each of the weights in these two sets are subtracted, squared, and aggregated together (like sum of squared errors) to obtain a single value, which denotes the amount of fine-tuning undergone by the model. We do the same for the ImageNet-trained model and compare the two. 
The FPUS23-trained model aggregates a value of $2.01\times10^{-7}$, whereas the ImageNet-trained model aggregates to $3.02\times10^{-6}$, which is more than $15\times$ larger. This implies that the former model requires less fine-tuning in comparison to the latter.  
Section~\ref{sec:Ablation} presents a comprehensive analysis of the knowledge retained and transferred by models trained on the \textit{FPUS23} dataset using ablation studies, followed by a discussion of the anatomy detection results for the trained DNN model. 


\subsection{Ablation Studies}
\label{sec:Ablation}

With no fine-tuning on the state-of-the-art dataset~\cite{burgos2020evaluation}, the FPUS23-trained model is able to make predictions with relatively higher accuracy when compared to an ImageNet-trained model, as shown by the results in Table~\ref{tab:errors}.

\begin{table}[t]
\centering 
{\fontsize{9pt}{9pt}\selectfont
\begin{tabular}{lcc}
\hline
Metric         & ImageNet-trained & FPUS23-trained \\ \hline
Number of Errors & 562 & 319 \\
Total Number of Images & 656 & 656 \\
Accuracy Percentage & 14.33 & 51.37 \\\hline
\end{tabular}}
\caption{Evaluation of the models trained using ImageNet and FPUS23 on the real-world fetal ultrasound dataset~\cite{burgos2020evaluation} without any fine-tuning.\label{tab:errors}}
\end{table}

Next, we provide a class-wise prediction breakdown of the models when fine-tuned on the state-of-the-art dataset. 
After ~15 epochs, the ImageNet-trained model converges to the same accuracy as the FPUS23-trained model, which requires just 1 epoch for fine-tuning. 
The results of these experiments are presented in Tables~\ref{tab:breakdown1} and \ref{tab:breakdown2}; the ImageNet-trained model is fine-tuned for 5 epochs and converges to the same accuracy-level as the FPUS23-trained model after 15 epochs.

We have also trained the two models using a 40\% subset of the real-world training data to achieve the same accuracy as the baseline (92\%) when using the FPUS23-trained model.
Whereas the ImageNet trained model is unable to achieve the same accuracy and falls short at 88\% accuracy. 

\begin{table}[h]
\centering
{\fontsize{9pt}{9pt}\selectfont
\begin{tabular}{cccccc}
                                                 &          & \multicolumn{4}{c}{Ground Truth} \\ \cline{3-6} 
\multicolumn{1}{l}{}                             &          & Abdomen & Brain & Femur & Thorax \\ \hline
\multicolumn{1}{c}{\multirow{4}{*}{Prediction}} & Abdomen  &   52   &   11  &   12  &   1    \\
\multicolumn{1}{c}{}                            & Brain    &   0    &   72  &   2   &   0    \\
\multicolumn{1}{c}{}                            & Femur    &   7    &   2   &   134 &   0    \\
\multicolumn{1}{c}{}                            & Thorax   &   13   &   9   &   5   &   336  \\ \hline
\multicolumn{1}{c}{}                            & Accuracy &  72.22 &  76.60& 87.58 & 99.70
\end{tabular}}
\caption{Class-wise prediction breakdown of the ImageNet-trained model when fine-tuned for the real-world dataset~\cite{burgos2020evaluation} after 5 epochs.\label{tab:breakdown1}}
\end{table}

\begin{table}[h]
\centering
{\fontsize{9pt}{9pt}\selectfont
\begin{tabular}{cccccc}
                                                 &          & \multicolumn{4}{c}{Ground Truth} \\ \cline{3-6} 
\multicolumn{1}{l}{}                             &          & Abdomen & Brain & Femur & Thorax \\ \hline
\multicolumn{1}{c}{\multirow{4}{*}{Prediction}} & Abdomen  &   39   &   0   &   5   &   0    \\
\multicolumn{1}{c}{}                            & Brain    &   3    &   91  &   7   &   6    \\
\multicolumn{1}{c}{}                            & Femur    &   21   &   1   &   137 &   0    \\
\multicolumn{1}{c}{}                            & Thorax   &   9    &   2   &   4   &   331  \\ \hline
\multicolumn{1}{c}{}                            & Accuracy &  54.17 &  96.81& 89.54 & 98.22
\end{tabular}}
\caption{Class-wise prediction breakdown of the FPUS23-trained model when fine-tuned for the real-world dataset~\cite{burgos2020evaluation} after just 1 epoch.\label{tab:breakdown2}}
\end{table}


Fig.~\ref{fig:ODEval} provides a sample of the anatomy detection results of the Faster-RCNN model with the ResNet34 backbone trained using the \textit{FPUS23} dataset.
As illustrated by the ground truth and prediction, in figs.~\ref{fig:ODEval}(a) and~\ref{fig:ODEval}(b), respectively, the precision achieved by the model is quite high, and can detect anatomies quite accurately and precisely in most instances, as illustrated by the overall mAP and mAR values of the model.
However, the model also misclassifies certain abstract artifacts during probe navigation as fetus anatomies, as illustrated by fig.~\ref{fig:ODEval}(c).
Fig.~\ref{fig:ODEval}(d) illustrates an instance wherein the model over- or under-estimates the bounds of the fetus anatomy.


\section{Conclusion and Future Work}
\label{sec:Conc}
In this paper, we present the \textit{FPUS23}, which is an ultrasound dataset of a fetus phantom at $23$ weeks gestation.
The data streams are collected and annotated by scientists with relevant fetal ultrasound experience to obtain information regarding the
\begin{inlinelist}
    \item diagnostic plane,
    \item fetus orientation,
    \item fetus anatomy, and
    \item their bounds, using box annotations.
\end{inlinelist}
The generated dataset is used to train a variety of deep learning models to illustrate the model's ability to extract vital information, which can be used to accurately distinguish among the classes in different categories and detect the fetus anatomy bounds.
Furthermore, to evaluate their deployability in portable resource-constrained devices, we evaluated the capability of a smaller DNN compressed using pruning and quantization to illustrate that smaller DNNs are equally competent at extracting relevant information from the dataset and are capable of execution on resource-constrained devices and embedded platforms.
The \textit{FPUS23} dataset is open-source and the trained models are accessible online.
In our future work, we plan to include annotated data of fetus phantoms at different gestation durations to offer a more comprehensive fetal ultrasound dataset.

\bibliographystyle{IEEEtran}
\bibliography{refs}

\begin{thebibliography}{10}
\providecommand{\url}[1]{#1}
\csname url@samestyle\endcsname
\providecommand{\newblock}{\relax}
\providecommand{\bibinfo}[2]{#2}
\providecommand{\BIBentrySTDinterwordspacing}{\spaceskip=0pt\relax}
\providecommand{\BIBentryALTinterwordstretchfactor}{4}
\providecommand{\BIBentryALTinterwordspacing}{\spaceskip=\fontdimen2\font plus
\BIBentryALTinterwordstretchfactor\fontdimen3\font minus
  \fontdimen4\font\relax}
\providecommand{\BIBforeignlanguage}[2]{{%
\expandafter\ifx\csname l@#1\endcsname\relax
\typeout{** WARNING: IEEEtran.bst: No hyphenation pattern has been}%
\typeout{** loaded for the language `#1'. Using the pattern for}%
\typeout{** the default language instead.}%
\else
\language=\csname l@#1\endcsname
\fi
#2}}
\providecommand{\BIBdecl}{\relax}
\BIBdecl

\bibitem{born2020pocovid}
J.~Born, G.~Br{\"a}ndle, M.~Cossio, M.~Disdier, J.~Goulet, J.~Roulin, and
  N.~Wiedemann, ``Pocovid-net: automatic detection of covid-19 from a new lung
  ultrasound imaging dataset (pocus),'' \emph{arXiv preprint arXiv:2004.12084},
  2020.

\bibitem{ebadi2021covidx}
A.~Ebadi, P.~Xi, A.~MacLean, S.~Tremblay, S.~Kohli, and A.~Wong,
  ``Covidx-us--an open-access benchmark dataset of ultrasound imaging data for
  ai-driven covid-19 analytics,'' \emph{arXiv preprint arXiv:2103.10003}, 2021.

\bibitem{shi2016stacked}
J.~Shi, S.~Zhou, X.~Liu, Q.~Zhang, M.~Lu, and T.~Wang, ``Stacked deep
  polynomial network based representation learning for tumor classification
  with small ultrasound image dataset,'' \emph{Neurocomputing}, vol. 194, pp.
  87--94, 2016.

\bibitem{EPIQ}
Philips, ``{EPIQ Elite},''
  \url{https://www.usa.philips.com/healthcare/product/HC795098/epiq-elite-a-new-class-of-premium-ultrasound-has-arrived}.

\bibitem{FetusPhantom}
{Kyoto Kagaku}, ``Fetus ultrasound examination phantom ``{SPACE FAN-ST}'',''
  \url{https://www.kyotokagaku.com/en/products_data/us-7_en/}, 2020.

\bibitem{al2020dataset}
W.~Al-Dhabyani, M.~Gomaa, H.~Khaled, and A.~Fahmy, ``Dataset of breast
  ultrasound images,'' \emph{Data in brief}, vol.~28, p. 104863, 2020.

\bibitem{leclerc2019deep}
S.~Leclerc, E.~Smistad, J.~Pedrosa, A.~{\O}stvik, F.~Cervenansky, F.~Espinosa,
  T.~Espeland, E.~A.~R. Berg, P.-M. Jodoin, T.~Grenier \emph{et~al.}, ``Deep
  learning for segmentation using an open large-scale dataset in 2d
  echocardiography,'' \emph{IEEE transactions on medical imaging}, vol.~38,
  no.~9, pp. 2198--2210, 2019.

\bibitem{valanarasu2020learning}
J.~M.~J. Valanarasu, R.~Yasarla, P.~Wang, I.~Hacihaliloglu, and V.~M. Patel,
  ``Learning to segment brain anatomy from 2d ultrasound with less data,''
  \emph{IEEE Journal of Selected Topics in Signal Processing}, vol.~14, no.~6,
  pp. 1221--1234, 2020.

\bibitem{patra2017learning}
A.~Patra, W.~Huang, and J.~A. Noble, ``Learning spatio-temporal aggregation for
  fetal heart analysis in ultrasound video,'' in \emph{Deep Learning in Medical
  Image Analysis and Multimodal Learning for Clinical Decision Support}.\hskip
  1em plus 0.5em minus 0.4em\relax Springer, 2017, pp. 276--284.

\bibitem{komatsu2021detection}
M.~Komatsu, A.~Sakai, R.~Komatsu, R.~Matsuoka, S.~Yasutomi, K.~Shozu, A.~Dozen,
  H.~Machino, H.~Hidaka, T.~Arakaki \emph{et~al.}, ``Detection of cardiac
  structural abnormalities in fetal ultrasound videos using deep learning,''
  \emph{Applied Sciences}, vol.~11, no.~1, p. 371, 2021.

\bibitem{qu2019deep}
R.~Qu, G.~Xu, C.~Ding, W.~Jia, and M.~Sun, ``Deep learning-based methodology
  for recognition of fetal brain standard scan planes in 2d ultrasound
  images,'' \emph{Ieee Access}, vol.~8, pp. 44\,443--44\,451, 2019.

\bibitem{van2019automated}
T.~L. van~den Heuvel, H.~Petros, S.~Santini, C.~L. de~Korte, and B.~van
  Ginneken, ``Automated fetal head detection and circumference estimation from
  free-hand ultrasound sweeps using deep learning in resource-limited
  countries,'' \emph{Ultrasound in medicine \& biology}, vol.~45, no.~3, pp.
  773--785, 2019.

\bibitem{sobhaninia2019fetal}
Z.~Sobhaninia, S.~Rafiei, A.~Emami, N.~Karimi, K.~Najarian, S.~Samavi, and
  S.~R. Soroushmehr, ``Fetal ultrasound image segmentation for measuring
  biometric parameters using multi-task deep learning,'' in \emph{2019 41st
  annual international conference of the IEEE engineering in medicine and
  biology society (EMBC)}.\hskip 1em plus 0.5em minus 0.4em\relax IEEE, 2019,
  pp. 6545--6548.

\bibitem{van2018automated}
T.~L. van~den Heuvel, D.~de~Bruijn, C.~L. de~Korte, and B.~v. Ginneken,
  ``Automated measurement of fetal head circumference using 2d ultrasound
  images,'' \emph{PloS one}, vol.~13, no.~8, p. e0200412, 2018.

\bibitem{burgos2020evaluation}
X.~P. Burgos-Artizzu, D.~Coronado-Guti{\'e}rrez, B.~Valenzuela-Alcaraz,
  E.~Bonet-Carne, E.~Eixarch, F.~Crispi, and E.~Gratac{\'o}s, ``Evaluation of
  deep convolutional neural networks for automatic classification of common
  maternal fetal ultrasound planes,'' \emph{Scientific Reports}, vol.~10,
  no.~1, pp. 1--12, 2020.

\bibitem{MatTrans}
Philips, ``X6-1: {xMATRIX} array transducer with purewave crystal technology,''
  \url{https://www.usa.philips.com/healthcare/product/HC989605409281/x6-1}.

\bibitem{salomon2011practice}
L.~J. Salomon, Z.~Alfirevic, V.~Berghella, C.~Bilardo, E.~Hernandez-Andrade,
  S.~Johnsen, K.~Kalache, K.-Y. Leung, G.~Malinger, H.~Munoz \emph{et~al.},
  ``Practice guidelines for performance of the routine mid-trimester fetal
  ultrasound scan,'' \emph{Ultrasound in Obstetrics \& Gynecology}, vol.~37,
  no.~1, pp. 116--126, 2011.

\bibitem{bethune2013pictorial}
M.~Bethune, E.~Alibrahim, B.~Davies, and E.~Yong, ``A pictorial guide for the
  second trimester ultrasound,'' \emph{Australasian journal of ultrasound in
  medicine}, vol.~16, no.~3, pp. 98--113, 2013.

\bibitem{boris_sekachev_2020_4009388}
\BIBentryALTinterwordspacing
B.~Sekachev, N.~Manovich, M.~Zhiltsov, A.~Zhavoronkov, D.~Kalinin, B.~Hoff,
  TOsmanov, D.~Kruchinin, A.~Zankevich, DmitriySidnev, M.~Markelov,
  Johannes222, M.~Chenuet, a~andre, telenachos, A.~Melnikov, J.~Kim, L.~Ilouz,
  N.~Glazov, Priya4607, R.~Tehrani, S.~Jeong, V.~Skubriev, S.~Yonekura, vugia
  truong, zliang7, lizhming, and T.~Truong, ``opencv/cvat: v1.1.0,'' Aug. 2020.
  [Online]. Available: \url{https://doi.org/10.5281/zenodo.4009388}
\BIBentrySTDinterwordspacing

\bibitem{he2016deep}
K.~He, X.~Zhang, S.~Ren, and J.~Sun, ``Deep residual learning for image
  recognition,'' in \emph{Proceedings of the IEEE conference on computer vision
  and pattern recognition}, 2016, pp. 770--778.

\bibitem{deng2009imagenet}
J.~Deng, W.~Dong, R.~Socher, L.-J. Li, K.~Li, and L.~Fei-Fei, ``Imagenet: A
  large-scale hierarchical image database,'' in \emph{2009 IEEE conference on
  computer vision and pattern recognition}.\hskip 1em plus 0.5em minus
  0.4em\relax Ieee, 2009, pp. 248--255.

\bibitem{ren2015faster}
S.~Ren, K.~He, R.~Girshick, and J.~Sun, ``Faster r-cnn: Towards real-time
  object detection with region proposal networks,'' \emph{Advances in neural
  information processing systems}, vol.~28, pp. 91--99, 2015.

\bibitem{elsken2019neural}
T.~Elsken, J.~H. Metzen, and F.~Hutter, ``Neural architecture search: A
  survey,'' \emph{The Journal of Machine Learning Research}, vol.~20, no.~1,
  pp. 1997--2017, 2019.

\bibitem{cheng2017survey}
Y.~Cheng, D.~Wang, P.~Zhou, and T.~Zhang, ``A survey of model compression and
  acceleration for deep neural networks,'' \emph{arXiv preprint
  arXiv:1710.09282}, 2017.

\bibitem{han2015deep}
S.~Han, H.~Mao, and W.~J. Dally, ``Deep compression: Compressing deep neural
  networks with pruning, trained quantization and huffman coding,'' \emph{arXiv
  preprint arXiv:1510.00149}, 2015.

\bibitem{khudia2021fbgemm}
D.~Khudia, J.~Huang, P.~Basu, S.~Deng, H.~Liu, J.~Park, and M.~Smelyanskiy,
  ``Fbgemm: Enabling high-performance low-precision deep learning inference,''
  \emph{arXiv preprint arXiv:2101.05615}, 2021.

\bibitem{sze2017efficient}
V.~Sze, Y.-H. Chen, T.-J. Yang, and J.~S. Emer, ``Efficient processing of deep
  neural networks: A tutorial and survey,'' \emph{Proceedings of the IEEE},
  vol. 105, no.~12, pp. 2295--2329, 2017.

\bibitem{marchisio2018prunet}
A.~Marchisio, M.~A. Hanif, M.~Martina, and M.~Shafique, ``Prunet: Class-blind
  pruning method for deep neural networks,'' in \emph{2018 International Joint
  Conference on Neural Networks (IJCNN)}.\hskip 1em plus 0.5em minus
  0.4em\relax IEEE, 2018, pp. 1--8.

\bibitem{fang2020densely}
J.~Fang, Y.~Sun, Q.~Zhang, Y.~Li, W.~Liu, and X.~Wang, ``Densely connected
  search space for more flexible neural architecture search,'' in
  \emph{Proceedings of the IEEE/CVF Conference on Computer Vision and Pattern
  Recognition}, 2020, pp. 10\,628--10\,637.

\bibitem{wang2020apq}
T.~Wang, K.~Wang, H.~Cai, J.~Lin, Z.~Liu, H.~Wang, Y.~Lin, and S.~Han, ``Apq:
  Joint search for network architecture, pruning and quantization policy,'' in
  \emph{Proceedings of the IEEE/CVF Conference on Computer Vision and Pattern
  Recognition}, 2020, pp. 2078--2087.

\end{thebibliography}
\newpage

\begin{IEEEbiography}[{\includegraphics[width=1in,height=1.25in,clip,keepaspectratio]{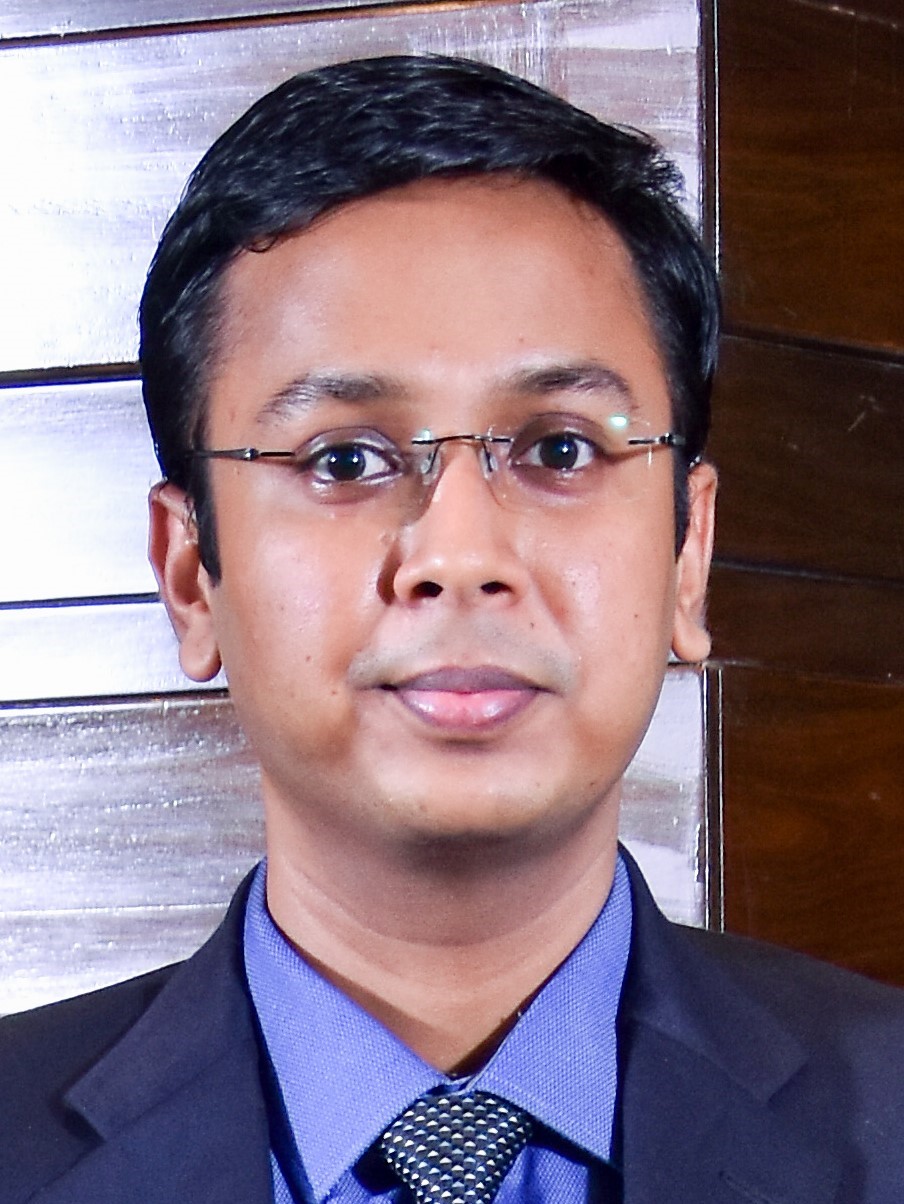}}]{Bharath Srinivas Prabakaran} is a PhD Student at the Doctoral College - Resilient Embedded Systems, Faculty of Informatics, TU Wien, Austria. He is working under the supervision of Prof. Muhammad Shafique. He graduated with a Bachelor of Engineering in Electrical and Electronics and a Master of Science in Biological Sciences from the Birla Institute of Technology and Science (BITS), Pilani, India in 2017. He was as a visiting researcher at TU Dresden, Germany for a span of 1 year from 2016, where he completed his master thesis. 
Since 2017, he worked as a project assistant at TU Wien for a span of 2 years before starting work on his doctoral dissertation.
His research interests include medical imaging, embedded machine learning, wearable architectures, healthcare systems, and energy-efficient technologies.
\end{IEEEbiography}

\begin{IEEEbiography}[{\includegraphics[width=1in,height=1.25in,clip,keepaspectratio]{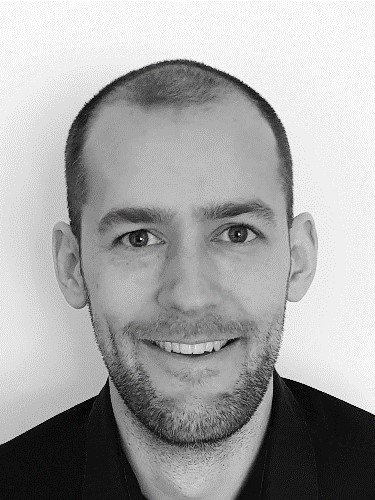}}]{Paul Hamelmann} received the M.Sc degree in biomedical engineering from the University of Twente, Enschede. 
In 2020, he obtained his Ph.D within the biomedical diagnostics lab (BM/d) of the Electrical Engineering faculty of the Eindhoven University of Technology (TU/e). 
He is currently working as a data scientist at Philips Research, Eindhoven. 
His current research interests include biomedical signal processing and (clinical) data science in the field of fetal and pregnancy monitoring.
\end{IEEEbiography}

\begin{IEEEbiography}[{\includegraphics[width=1in,height=1.25in,clip,keepaspectratio]{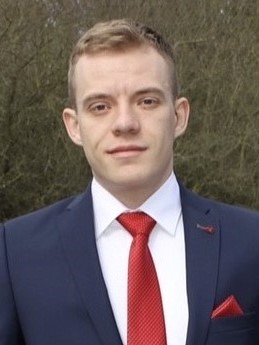}}]{Erik Ostrowski} (S’23) is a student at the PhD School of Informatics, TU Wien, Austria. He graduated with a Bachelor and a Master of Science in business mathematics from the Universität zu Köln, Cologne, Germany in 2020. His research interests include Computer Vision, Semantic Segmentation, Weak Supervision, Healthcare Systems, Energy-efficient Technologies, AI \& Machine Learning.
\end{IEEEbiography}

\begin{IEEEbiography}[{\includegraphics[width=1in,height=1.25in,clip,keepaspectratio]{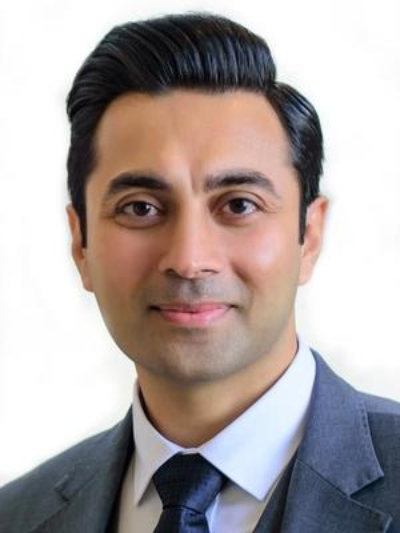}}]{Muhammad Shafique} (M’11 - SM’16) received the Ph.D. degree in computer science from the Karlsruhe Institute of Technology (KIT), Germany, in 2011. Afterwards, he established and led a highly recognized research group at KIT for several years as well as conducted impactful collaborative R\&D activities across the globe. In Oct.2016, he joined the Institute of Computer Engineering at the Faculty of Informatics, Technische Universit{\"a}t Wien (TU Wien), Vienna, Austria as a Full Professor of Computer Architecture and Robust, Energy-Efficient Technologies. Since Sep.2020, Dr. Shafique is with the New York University (NYU), where he is currently a Full Professor and the director of eBrain Lab at the NYU-Abu Dhabi in UAE, and a Global Network Professor at the Tandon School of Engineering, NYU-New York City in USA. He is also a Co-PI/Investigator in multiple NYUAD Centers, including Center of Artificial Intelligence and Robotics (CAIR), Center of Cyber Security (CCS), Center for InTeractIng urban nEtworkS (CITIES), and Center for Quantum and Topological Systems (CQTS).

His research interests are in AI \& machine learning hardware and system-level design, brain-inspired computing, machine learning security and privacy, quantum machine learning, cognitive autonomous systems, wearable healthcare, energy-efficient systems, robust computing, hardware security, emerging technologies, FPGAs, MPSoCs, and embedded systems. His research has a special focus on cross-layer analysis, modeling, design, and optimization of computing and memory systems. The researched technologies and tools are deployed in application use cases from Internet-of-Things (IoT), Smart Cyber-Physical Systems (CPS), and ICT for Development (ICT4D) domains. Dr. Shafique has given several Keynotes, Invited Talks, and Tutorials, as well as organized many special sessions at premier venues. He has served as the PC Chair, General Chair, Track Chair, and PC member for several prestigious IEEE/ACM conferences. Dr. Shafique holds one U.S. patent, and has (co-)authored 6 Books, 10+ Book Chapters, 350+ papers in premier journals and conferences, and 100+ archive articles. He received the 2015 ACM/SIGDA Outstanding New Faculty Award, the AI 2000 Chip Technology Most Influential Scholar Award in 2020 and 2022, the ASPIRE AARE Research Excellence Award in 2021, six gold medals, and several best paper awards and nominations at prestigious conferences. He is a senior member of the IEEE and IEEE Signal Processing Society (SPS), and a member of the ACM, SIGARCH, SIGDA, SIGBED, and HIPEAC.
\end{IEEEbiography}

\EOD

\end{document}